\documentclass[prb,twocolumn,nobalancelastpage,amsmath,amssymb]{revtex4}
\usepackage{graphicx}
\usepackage{bm}
\usepackage{bbm}
\usepackage{amssymb}
\begin{document}
\date{\today}

\title{
Chiral current - phase relation of topological Josephson junctions: \\
A signature of the $4\pi$ - periodic Josephson effect
}

\author{G. Tkachov}

\affiliation{
Institute of Physics, Augsburg University, 86135 Augsburg, Germany}

\begin{abstract}
The $4\pi$ - periodic Josephson effect is an indicator of Majorana zero modes and a ground-state degeneracy 
which are central to topological quantum computation.
However, the observability of a $4\pi$ - periodic Josephson current-phase relation (CPR) is hindered by the necessity to fix the fermionic parity.
As an alternative to a $4\pi$ - periodic CPR, this paper proposes a chiral CPR for the $4\pi$ - periodic Josephson effect.
This is a CPR of the form $J(\phi) \propto C \, |\sin(\phi/2)|$, describing a unidirectional supercurrent with the chirality $C= \pm 1$. 
Its non-analytic dependence on the Josephson phase difference $\phi$ translates into the $4\pi$ - periodic CPR 
$J(\phi) \propto \sin(\phi/2)$. The proposal requires a spin-polarized topological Josephson junction  
which is modeled here as a short link between spin-split superconducting channels at the edge of a two-dimensional topological insulator.
In this case, $C$ coincides with the Chern number of the occupied spin band of the topological insulator. 
The paper details three scenarios of achieving a chiral CPR: by Zeeman-like splitting only, by Zeeman splitting combined with bias currents, 
and by an external out-of-plane magnetic field.
\end{abstract}

\maketitle

\section{introduction}

In topological superconductors, electron pairs condense into a collective gapped state that coexists with gapless Majorana fermions on defects, 
harboring many unconventional properties. \cite{Kitaev01,Nayak08,Tanaka12,Sato17,Haim18}
Illustrative examples are one-dimension (1D) $p$-wave superconductors \cite{Kitaev01} and superconductor/semiconductor wires 
(see, e.g., Refs.\onlinecite{Sau10,Alicea10,Lutchyn10,Oreg10}) whose boundaries host a pair of Majorana zero modes (MZMs). 
Such 1D systems possess two ground states related by a permutation of the Majorana degrees of freedom 
and corresponding to two (even and odd) fermionic parities, 
which offers a platform for topological quantum computation.\cite{Kitaev01,Beenakker13a,Das Sarma15}

A striking manifestation of topological superconductivity occurs in Josephson junctions (JJs) of two Majorana wires brought into electric contact.
A  change of the Josephson phase difference by $2\pi$ effectively causes swapping the MZMs and a transition between the ground states.
This implies the $4\pi$ - periodicity of superconducting properties, as another phase advance of $2\pi$ is needed 
to recover the same ground state.\cite{Kitaev01}   
First proposed for model $p$-wave superconductors,\cite{Kitaev01,Kwon04}
such $4\pi$ - periodic topological superconductivity is also expected in hybrid structures of conventional superconductors and 
spin-orbit-coupled normal materials, which has been causing a surge of interest in this and related phenomena,
both in theory (see, e.g., Refs. \onlinecite{Fu08,Fu09,Tanaka09,Qi10,Badiane11,Dominguez12,San-Jose12,Pikulin12,GT13,Beenakker13b,Badiane13,Peng16}) 
and in experiment (see, e.g., Refs. \onlinecite{Rokhinson12,Wiedenmann16,Laroche17,Kayyalha19}).

Most of the recent research on the $4\pi$ - periodic Josephson effect 
has been dealing with or implies out-of-equilibrium AC properties of JJs under external driving.
If the current-carrying states have equilibrium occupations, 
the resulting periodicity of the Josephson current-phase relation (CPR) is $2\pi$,
i.e. the same as in non-topological JJs, unless the fermionic parity is constrained. \cite{Beenakker13b} 
Still, the ability to access the $4\pi$ - periodic Josephson effect through an equilibrium CPR despite its conventional periodicity 
is beneficial, as such CPRs are the most common characteristics of JJs, 
and there exist well established techniques for their measurement. \cite{Van Harlingen95,Tsuei00,Golubov04}
Furthermore, beside the periodicity of the CPR, there are other indicators of the $4\pi$ - periodic Josephson effect at equilibrium,
such as magnetic oscillations of the critical current with the doubled period $2\Phi_0 =h/e$ 
in the magnetic flux enclosed in the JJ. \cite{GT19a} 
A different type of the magnetic-field behavior has been predicted for semiconductor nanowire JJs, \cite{Cayao17}
where magnetic oscillations of the critical current indicate the splitting of the MZMs in finite-length wires.

This paper takes a closer look at the CPR of a topological JJ,
aiming to identify the change of its ground state upon an adiabatic phase advance.
We consider a short JJ at the edge of a 2D topological insulator (2DTI) with a uniform Zeeman-like spin splitting. 
A related model was used earlier in Refs. \onlinecite{Dolcini15} and \onlinecite{GT17a} in the context of magnetoelectric phenomena in quantum spin-Hall insulators. 
Unlike those works, here we focus on spin-polarized $4\pi$ - periodic ground states and topological transitions between them. 
Our goal is to demonstrate that the CPR becomes chiral in the sense that 
the two spin-polarized ground states carry the current in the same direction, 
and that such an anomalous CPR reveals the $4\pi$ - periodic Josephson effect.
The following sections explain the details of the calculations and provide an extended discussion of the results.

\section{Topological Josephson junction. Model}
\label{Model}

\begin{figure}[t]
\begin{center}
\includegraphics[width=85mm]{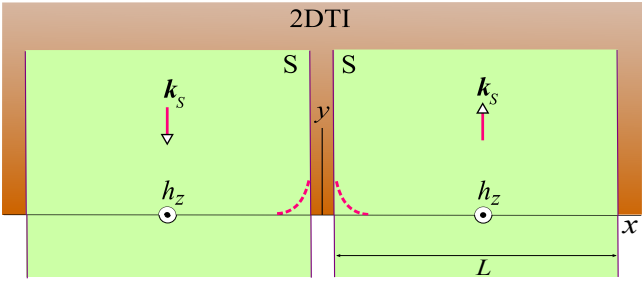}
\end{center}
\caption{
Schematic of a topological JJ created by placing two superconducting strips across the edge of a 2DTI.
}
\label{TJJ}
\end{figure}

The system consists of a 2DTI and two superconducting strips placed over its edge (see Fig. \ref{TJJ}). 
The edge region between the superconducting contacts acts as a Josephson weak link that 
can be modeled by an effective 1D Bogoliubov-de Gennes (BdG) Hamiltonian: 

\begin{eqnarray}
{\cal H} =
\left[%
\begin{array}{cc}
\upsilon s_z p_x - \mu +  h_z s_z  & \Delta(x) \\
 \Delta^*(x)  & -(\upsilon s_z p_x - \mu) + h_z s_z
\end{array}
\right].
\label{H}
\end{eqnarray}
Here, $\Delta(x)$ is the pair potential; 
the normal edge Hamiltonian consists of the kinetic energy $\upsilon s_z p_x - \mu$ and the spin-splitting potential $h_z s_z$,
where $\upsilon$, $s_z$, $p_x=-i\hbar\partial_x$, $\mu$, and $h_z$ are 
the edge-state velocity, spin Pauli matrix, momentum operator, chemical potential, and the spin-splitting energy, respectively. 
The origin of the spin splitting depends on the context, e.g., the paramagnetism of a magnetically doped 2DTI, 
the Zeeman effect of an out-of-plane magnetic field or the magnetoelectric effect 
caused by the spin-momentum locking. \cite{GT17a} 

For a half-space 2DTI, the pair potential $\Delta(x)$ can be written as \cite{GT19a} 

\begin{eqnarray}
\Delta(x) = \frac{\int^\infty_0 \Delta_{_{2D}}(x,y) f(y)dy}{\int^\infty_0 f(y)dy} ,
\label{Delta}
\end{eqnarray}
where $\Delta_{_{2D}}(x,y)$ is the proximity-induced pair potential underneath a superconducting strip, 
and $f(y)$ is the transverse wave function of the edge state.  
Equation (\ref{Delta}) is thus a weighted average of $\Delta_{_{2D}}(x,y)$ in the half space $y > 0$
with the weight $f(y)$. We adopt the Bernevig-Hughes-Zhang model \cite{Bernevig06,Liu16} in which the edge wave function is given by
\begin{equation}
f(y)= e^{ -\varkappa_+ y } - e^{ -\varkappa_- y }, \quad
\varkappa_\pm=\frac{|{\cal A}|}{2|{\cal B}|} \pm \sqrt{ \frac{{\cal A}^2}{4{\cal B}^2} + \frac{ {\cal M} }{ {\cal B} } },
\label{f}
\end{equation}
where $\varkappa_\pm$ are the decay constants depending on the band structure parameters ${\cal A}, {\cal B}$, and $M$. 
Here, ${\cal A}$ is the strength of the spin - momentum locking, while ${\cal M}$ and ${\cal B}$ are the bulk band gap and its curvature, respectively.
We note that the edge-state velocity $\upsilon$ in Eq. (\ref{H}) can be expressed in terms of the parameters ${\cal A}$ and ${\cal M}$ as~\cite{GT19a} 
\begin{equation}
\upsilon = |{\cal A}|{\rm sgn}(M)/\hbar. 
\label{v}
\end{equation}
The sign of the band gap, ${\rm sgn}(M)$, is directly related to the Chern number of a bulk electronic band for given spin orientation.~\cite{Bernevig06,Liu16} 

Equation (\ref{Delta}) is applicable for an inhomogeneous superconducting order parameter, 
which is for example the case in the presence of external magnetic fields or bias currents.  
In such cases, the phase of the order parameter varies in space, so $\Delta_{_{2D}}(x,y)$ is an oscillating function.  
We assume that the phase has a uniform gradient in the transverse ($y$) direction with opposite signs in  
the superconducting leads, as depicted in Fig. \ref{TJJ}. 
Such a situation can be realized either by passing antiparallel bias currents in the leads or by applying an out-of-plane magnetic field.
This provides means to tune the topological superconductivity in JJs (see Secs. \ref{h+kS} and \ref{B} for a detailed discussion).
 
In a given contact (say, in the right one), the pair potential can be written as $\Delta_{_{2D}}(y)= \Delta_0 e^{ i\varphi(y) }$, 
with a constant amplitude $\Delta_0$ and the nonuniform phase  
\begin{eqnarray}
\varphi(y) = \varphi_0 + k_{_S}y.
\label{phase_right}
\end{eqnarray}
In the case of the bias current, we assume that
the phase gradient $k_{_S} = \partial_y\varphi$ is generated by a uniform current density $j= \rho_{_S} \partial_y\varphi$ 
in the overlying superconducting strip, so that $k_{_S} =j/\rho_{_S}$ (where $\rho_{_S}$ is the superfluid stiffness). 
The phase $\varphi(y)$ is counted from its value at the edge, $\varphi_0$.
In the other contact, the pair potential has a similar form, ${\bar \Delta}_{_{2D}}(y)= \Delta_0 e^{ i{\bar \varphi}(y) }$, with
\begin{eqnarray}
{\bar \varphi}(y) = {\bar \varphi}_0 - k_{_S}y,
\label{phase_left}
\end{eqnarray}
only the current direction and the constant ${\bar \varphi}_0$ are different.

From Eqs. (\ref{Delta}) - (\ref{phase_left}) we readily obtain the edge pair potentials in the right and left contacts as

\begin{eqnarray}
\Delta = 
|\Delta| e^{ i(\varphi_0 + \vartheta)}  
\,\,\,
{\rm (right)},
\qquad
{\bar \Delta} = 
|\Delta| e^{ i({\bar \varphi}_0 - \vartheta)} 
\,\,\,
{\rm (left)},
\label{Delta_RL}
\end{eqnarray}
where the modulus $|\Delta|$ and the phase $\vartheta$ both depend on the phase gradient: \cite{GT19a}

\begin{eqnarray}
|\Delta(k_{_S})| &=&  \Delta_0 \frac
{\varkappa_+\varkappa_-}
{\sqrt{
(\varkappa_+\varkappa_- - k_{_S}^2)^2 + (\varkappa_+ + \varkappa_-)^2 k_{_S}^2
}
},
\label{Delta_mod}\\
\vartheta(k_{_S}) &=& \arctan \frac{ (\varkappa_+ + \varkappa_-)k_{_S} }{\varkappa_+\varkappa_- - k_{_S}^2}.
\label{theta}
\end{eqnarray}
The function $|\Delta(k_{_S})|$ accounts for a partial reduction of the proximity-induced energy gap by the supercurrent.
This is due to the averaging of the spatial oscillations of the order parameter superimposed on the exponential decay of the edge state.
The intrinsic phase $\vartheta(k_{_S})$ is another consequence of the complexity of the order parameter that comes out of the averaging 
in Eq. (\ref{Delta}). Generally, in noncentrosymmetric superconductors the phase gradient induces also a non-unitary triplet order parameter. \cite{GT17b,GT19b} 
This aspect of the problem will be discussed elsewhere.

To sum up, we cast the BdG Hamiltonian (\ref{H}) as 
\begin{equation}
{\cal H}(x) =  h_z s_z +  \tau_3 \upsilon s_z p_x  + \tau_1 \Delta_{_{Re}}(\phi) - \tau_2 \Delta_{_{Im}}(\phi) {\rm sgn}(x),
\label{H_Nambu}
\end{equation}
where  $\tau_1, \tau_2$, and $\tau_3$ are the Nambu-Pauli matrices, 
$\Delta_{_{Re}}(\phi)$ and $\Delta_{_{Im}}(\phi)$ are the real and imaginary parts of the pair potential

\begin{eqnarray}
\Delta_{_{Re}}(\phi) &=& |\Delta| \cos\left(\vartheta + \phi/2 \right),
\label{Delta_Re}\\
\Delta_{_{Im}}(\phi) &=& |\Delta| \sin\left(\vartheta + \phi/2 \right),
\label{Delta_Im}
\end{eqnarray}
and $\phi = \varphi_0 - {\bar \varphi}_0$ is the external phase bias 
[the average phase $(\varphi_0 + {\bar \varphi}_0)/2$ and the chemical potential have both been gauged out].
In Eq. (\ref{H_Nambu}), the link is reduced to a single point ($x=0$), which is the main approximation for JJs 
with the normal spacer much shorter than the superconducting coherence length.

\section{Ground-state doublet. Topological transmutations and $4\pi$ periodicity}

In this section, we examine the transformation properties of the JJ ground state under an adiabatic phase change from $\phi$ to $\phi + 2\pi$.
In terms of the BdG wave function, this transformation can be expressed as 

\begin{equation}
\Psi(x, \phi) \longrightarrow \tau_3 \Psi(x, \phi + 2\pi),
\label{Psi_2pi}
\end{equation}
leaving the Hamiltonian invariant:

\begin{equation}
\tau_3 {\cal H}(x, \phi + 2\pi) \tau_3 = {\cal H}(x, \phi).
\label{H_2pi}
\end{equation}
That is, the states $\tau_3 \Psi(x, \phi + 2\pi)$ and $\Psi(x, \phi)$ both correspond to the same energy.
In non-topological JJs, these two states are just identical, meaning no degeneracy associated with the phase translation 
$\phi \to \phi + 2\pi$. However, in topological edge JJs, the transformation (\ref{Psi_2pi}) does produce a new orthogonal ground state, 
implying a ground-state doublet similar to Kitaev's model. \cite{Kitaev01}
The BdG formalism is quite different from that of Kitaev's model. 
To draw parallels between them, 
we map the BdG Hamiltonian to the Jackiw-Rebbi model describing a paradigmatic topologically nontrivial 1D fermion system. \cite{Jackiw76} 
The map employs a unitary transformation similar to that in Ref. \onlinecite{GT19a}.
The analysis below generalizes the approach of Ref. \onlinecite{GT19a} to account for the spin splitting in Eq. (\ref{H_Nambu}).

\subsection{Mapping to Jackiw-Rebbi model}
\label{JR}

The idea is to make a time-dependent unitary transformation,

\begin{equation}
\Psi(x,t) = U(t)  \, \Psi^\prime(x,t), \quad  U(t) = e^{-i[ \tau_1 \Delta_{_{Re}}(\phi) + s_z h_z ] t/\hbar},
\label{U}
\end{equation}
of the BdG equation $i\hbar\partial_t \Psi(x,t) = {\cal H}(x) \Psi(x,t)$, 
bringing the BdG Hamiltonian to the form

\begin{eqnarray}
{\cal H}^\prime(x,t) 
=  U^\dagger(t)  {\cal H}(x) U(t) - i\hbar U^\dagger(t) \partial_t U(t)
\label{H^prime_1}
\end{eqnarray}
\begin{eqnarray}
=  U^\dagger(t)  [\tau_3 \upsilon s_z p_x - \tau_2 \Delta_{_{Im}}(\phi) {\rm sgn}(x)] U(t)
\label{H^prime_2}
\end{eqnarray}
\begin{eqnarray}
=   e^{2i \tau_1 \Delta_{_{Re}}(\phi) t/\hbar} [\tau_3 \upsilon s_z p_x - \tau_2 \Delta_{_{Im}}(\phi) {\rm sgn}(x)].
\label{H^prime_3}
\end{eqnarray}
Up to the first (time-dependent) factor, the new Hamiltonian (\ref{H^prime_3}) is analogous to that of the Jackiw-Rebbi model 
for the 1D Dirac fermion in a soliton background. 
The imaginary part of the pair potential $\Delta_{_{Im}}(\phi){\rm sgn}(x)$ acts as a sharp Jackiw-Rebbi soliton,
while the chiral symmetry $\tau_1 {\cal H}^\prime(x,t) \tau_1 = - {\cal H}^\prime(x,t)$ ensures the existence of MZMs. 
The latter are the eigenstates of the chirality operator $\tau_1$, 
satisfying the equation ${\cal H}^\prime(x,t) \Psi^\prime(x,t)=0$ which in fact is time-independent:

\begin{equation}
[\hbar\upsilon\partial_x - \tau s \Delta_{_{Im}}(\phi) {\rm sgn}(x)]\Psi^\prime(x)=0,
\label{Equation_MZMs}
\end{equation}
where $\tau$ and $s$ are the eigenvalues of $\tau_1$ and $s_z$, respectively. 
The solutions of Eq. (\ref{Equation_MZMs}) characterize possible ground states of the model.

\subsection{Majorana zero modes and ground-state transmutations}
\label{MZMs}

Equation (\ref{Equation_MZMs}) has MZM solutions at $x=0$.
The wave function is evanescent, $\Psi^{^{MZM}}(x) \propto e^{-k |x|}$, where the inverse decay length is given by 
$k = - \tau s \Delta_{_{Im}}(\phi)/(\hbar \upsilon)$. 
The normalizability condition $k > 0$ imposes the constraint on the quantum numbers \cite{GT19a}

\begin{eqnarray}
\tau s =  - {\rm sgn}(\upsilon \Delta_{_{Im}}(\phi)) = - {\rm sgn}({\cal M} \sin(\vartheta + \phi/2)),
\label{Sign}
\end{eqnarray}
see also Eqs. (\ref{v}) and (\ref{Delta_Im}).
Since the product of the eigenvalues $\tau$ and $s$ is fixed, there are only two different quantum numbers 
and, consequently, two orthogonal MZMs. It is convenient to label them by the eigenvalues $\tau = \pm 1$.
Then, the two states have opposite spin orientations 

\begin{equation}
s = \pm \sigma(\phi), \qquad \sigma(\phi)= - {\rm sgn}({\cal M} \sin(\vartheta + \phi/2)),
\label{Sigma}
\end{equation}
and their wave functions are given, up to a normalizing factor, by

\begin{equation}
\Psi^{^{MZM}}_+(x, \phi) =
\left[
\begin{array}{c}
1 \\ 1
\end{array}
\right]
\otimes 
\left[
\begin{array}{c}
\frac{1 + \sigma(\phi)}{2}  \\  \frac{1 - \sigma(\phi)}{2} 
\end{array}
\right]
e^{- \left| \frac{ \Delta_{_{Im}}(\phi) }{\hbar\upsilon} x \right| },
\label{MZM_+}
\end{equation}
and 

\begin{equation}
\Psi^{^{MZM}}_-(x, \phi) =
\left[
\begin{array}{c}
1 \\ -1
\end{array}
\right]
\otimes 
\left[
\begin{array}{c}
\frac{1 - \sigma(\phi)}{2}  \\  \frac{1 + \sigma(\phi)}{2} 
\end{array}
\right]
e^{- \left| \frac{ \Delta_{_{Im}}(\phi) }{\hbar\upsilon} x \right| },
\label{MZM_-}
\end{equation}
where $\otimes$ means the direct product of the Nambu and spin states.

Let us look at the phase dependence of Eq. (\ref{MZM_+}). 
With an adiabatic phase change from $\phi$ to $\phi + 2\pi$, the singular function $\sigma(\phi)$  in Eq. (\ref{Sigma}) 
flips its sign, describing the transmutation of the spin state 

\begin{equation}
\left[
\begin{array}{c}
\frac{1 + \sigma(\phi)}{2}  \\  \frac{1 - \sigma(\phi)}{2} 
\end{array}
\right]
\longrightarrow 
\left[
\begin{array}{c}
\frac{1 - \sigma(\phi)}{2}  \\  \frac{1 + \sigma(\phi)}{2} 
\end{array}
\right].
\label{Spin_trans}
\end{equation}
This is a topological transition during which the mass term $\Delta_{_{Im}}(\phi)$ in Eq. (\ref{Equation_MZMs}) passes through zero, 
while a kink phase profile at the junction switches to an anti-kink one. 
The reversal of the spin state in Eq. (\ref{Spin_trans}) is consistent with the Jackiw-Rebbi model 
where a kink and an anti-kink host zero modes with the opposite spin projections. 
Up to the unitary rotation with $\tau_3$, the new state coincides with that in Eq. (\ref{MZM_-}): 

\begin{equation}
\tau_3 \Psi^{^{MZM}}_+(x, \phi + 2\pi) =  \Psi^{^{MZM}}_-(x, \phi).
\label{Trans_2pi}
\end{equation}
Repeating the same transformation yields

\begin{equation}
\Psi^{^{MZM}}_\pm(x, \phi + 4\pi) =  \Psi^{^{MZM}}_\pm(x, \phi).
\label{Trans_4pi}
\end{equation}
That is, there are two orthogonal $4\pi$ - periodic ground states 
transforming into each other upon an adiabatic $2\pi$ phase advance 
in a similar manner as the ground states of the lattice model of Ref.  \onlinecite{Kitaev01}.
However, here, the superconductivity is time-reversal invariant (in the absence of external fields and bias currents)
due to the spin-momentum locking in the 2DTI. 
Consequently, the two ground states carry opposite spin polarizations, 
producing no net magnetization for $h_z =0$.
It is worth emphasizing that, while the phase evolution is adiabatic, 
the transition between the ground states is discontinuous, 
which is described by the singular function $\sigma(\phi)$  in Eq. (\ref{Sigma}).

\subsection{$4\pi$ - periodic Andreev bound states}
\label{ABSs}

The MZMs of the transformed Hamiltonian (\ref{H^prime_3}) are related to the ABSs of the original Hamiltonian (\ref{H_Nambu}). 
The ABSs are composed of the MZMs residing at the adjacent ends of the right and left superconductors.
Here, we construct the $4\pi$ - periodic ABSs, using the unitary map (\ref{U}) to the Jackiw-Rebbi model.
We just need to replace $\Psi^\prime(x,t)$ in Eq. (\ref{U}) with the stationary MZM solutions (\ref{MZM_+}) and (\ref{MZM_-}). 
That is, the ABSs come as the unitary time evolution of the stationary MZMs:

\begin{eqnarray}
\Psi^{^{ABS}}(x,t) 
&=& U(t) \Psi^{^{MZM}}_\tau (x)  
\label{ABS_MZM_1}\\
&=& e^{-i[ \tau_1 \Delta_{_{Re}}(\phi) + s_z h_z ] t/\hbar} \Psi^{^{MZM}}_\tau (x).
\label{ABS_MZM_2}
\end{eqnarray}
Clearly, the ABSs carry the same quantum numbers as the MZMs [see Eq. (\ref{Sigma})],
so the above relation is reduced to the multiplication by a time-dependent phase factor:

\begin{equation}
\Psi^{^{ABS}}_\tau(x,t) =  e^{-i \tau [  \Delta_{_{Re}}(\phi) + h_z \sigma(\phi) ] t/\hbar} \Psi^{^{MZM}}_\tau (x),
\label{ABS_MZM_3}
\end{equation}
from which we identify the ABS energy levels as

\begin{equation}
E_\pm (\phi) = \pm \bigl[ 
|\Delta| \cos\left(\vartheta + \phi/2 \right) + h_z  \sigma(\phi)
\bigr], 
\label{E_4pi}
\end{equation}
where the sign $\pm$ refers to the eigenvalue $\tau$ of the chirality matrix $\tau_1$. 
The ABS levels inherit the transmutation property of the MZMs [cf. Eq. (\ref{Trans_2pi})],

\begin{equation}
E_+(\phi + 2\pi) = E_-(\phi).
\label{ABS_trans}
\end{equation}
This, again, indicates the topological degeneracy associated with a $2\pi$ phase translation. 
Also, while evolving from $E_+(\phi)$ to $E_+(\phi + 2\pi)$, the level passes through zero, 
so the new energetically favorable ground state should have 
a different fermionic parity \cite{Kitaev01} compared to the state hosting $E_+(\phi)$.

\section{Chiral current-phase relation}

\subsection{CPR. Preliminaries}
\label{CPR}

In short JJs, a major contribution to the phase-dependent supercurrent comes from the ABSs. \cite{Golubov04}
To obtain the CPR, we use the thermodynamic relation between the supercurrent $J(\phi)$ and the ABS levels,

\begin{equation}
J(\phi) = 
\frac{e}{\hbar} \frac{\partial E_+ (\phi)}{\partial \phi} n[E_+ (\phi)] + \frac{e}{\hbar} \frac{\partial E_- (\phi)}{\partial \phi} n[E_- (\phi)],
\label{J_short}
\end{equation}
where $n[E_\pm (\phi)]$ is the Fermi occupation number. 
We will focus on the zero-temperature case in which the ABS levels are occupied according to $n[E_\pm (\phi)] = \frac{1}{2} ( 1 - {\rm sgn}[E_\pm (\phi)]  )$.

Each contribution in Eq. (\ref{J_short}) is a $4\pi$ - periodic Josephson current. 
However, the topological degeneracy [see Eq. (\ref{ABS_trans})] makes the net current $2\pi$  periodic, 
as the two contributions simply swap upon a $2\pi$ phase advance.
Nevertheless, the CPR (\ref{J_short}) is a characteristic of the $4\pi$ - periodic Josephson effect
since the current is carried by the doublet of the $4\pi$ - periodic ground states.
The latter are represented by the occupied branches of the ABS levels.
In this paper, the notion of the $4\pi$ - periodic Josephson effect refers not to the periodicity of the CPR,
but to the topological properties of the current-carrying states. 

At zero temperature, Eq. (\ref{J_short}) reads

\begin{equation}
J(\phi) = -\frac{e}{\hbar} \frac{\partial E_+ (\phi)}{\partial\phi} {\rm sgn}[E_+ (\phi)].
\label{J_short_T0}
\end{equation}
We note that the derivative of the ABS level $E_+ (\phi)$ (\ref{E_4pi}) 
is continuous at the singularities of $\sigma(\phi)$, hence

\begin{equation}
J(\phi) =  
\frac{e |\Delta|}{2\hbar}
\sin\left(\vartheta + \frac{\phi}{2} \right) 
{\rm sgn} 
\Biggl[
\cos\left(\vartheta + \frac{\phi}{2}\right) + \frac{h_z}{ |\Delta|} \sigma(\phi)
\Biggr].
\label{J}
\end{equation}
Generally, the shape of the CPR $J(\phi)$ depends on both the spin splitting energy $h_z$ and the phase gradient $k_{_S}$. 
The latter enters through the modulus $|\Delta|$ (\ref{Delta_mod}) and the phase $\theta$ (\ref{theta}).

\subsection{Case $h_z \not =0$ and $k_{_S}=0$}
\label{h}

It is instructive to discuss first the effect of the spin splitting $h_z$ only, setting 

\begin{equation}
k_{_S}=0, \qquad |\Delta| = \Delta_0, \qquad \theta =0
\label{Parameters_kS0}
\end{equation}
in Eqs. (\ref{E_4pi}) and (\ref{J}). The corresponding ABSs levels and CPR are then given by

\begin{equation}
E_\pm (\phi) = \pm \bigl[ 
\Delta_0 \cos\left(\phi/2 \right) - h_z  {\rm sgn}({\cal M} \sin(\phi/2))
\bigr], 
\label{E_h}
\end{equation}

\begin{equation}
J(\phi) =  
\frac{e \Delta_0}{2\hbar} \sin\left(\frac{\phi}{2}\right) 
{\rm sgn} \Biggl[ 
\cos\frac{\phi}{2}  - \frac{h_z}{\Delta_0}{\rm sgn}\left({\cal M} \sin\frac{\phi}{2}\right) 
\Biggr] .
\label{J_h}
\end{equation}

\begin{figure}[t]
\begin{center}
\includegraphics[width=75mm]{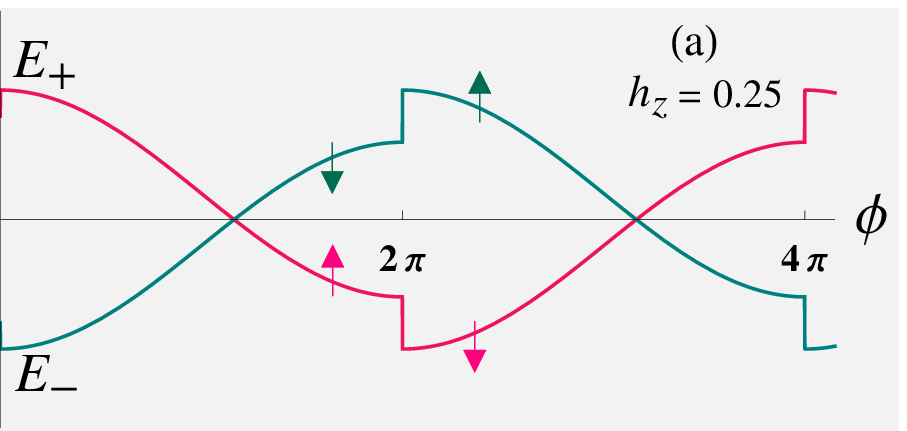}
\vskip 0.07cm
\includegraphics[width=75mm]{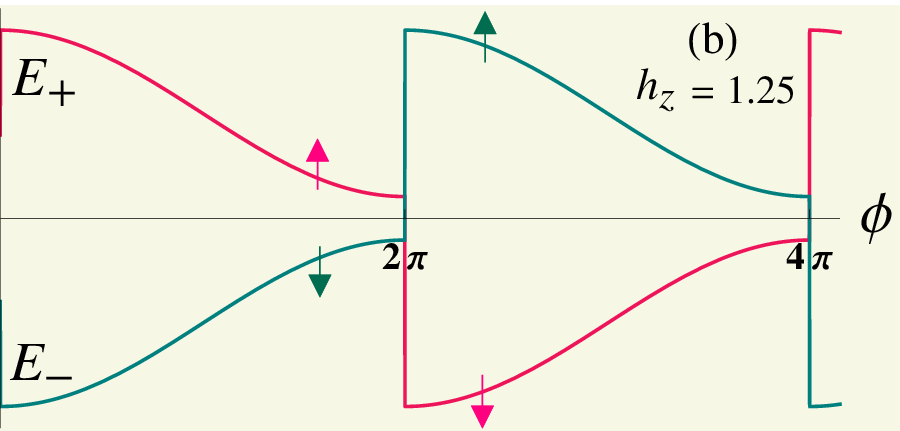}
\vskip 0.07cm
\includegraphics[width=75mm]{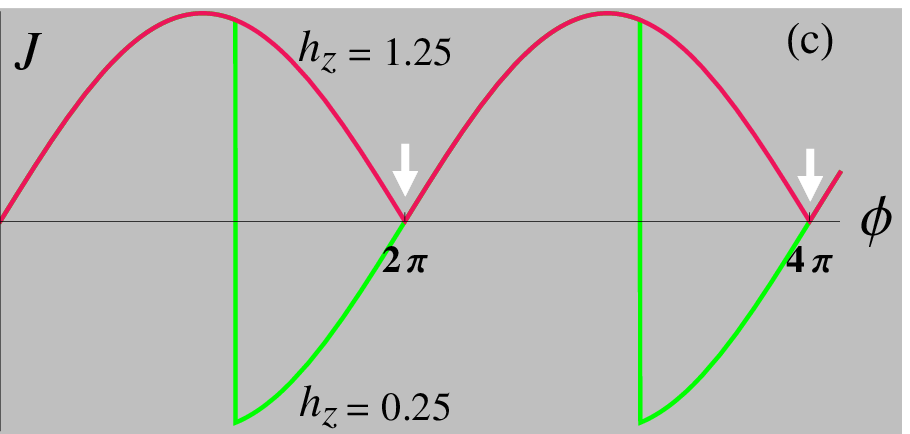}
\end{center}
\caption{
ABS levels $E_\pm(\phi)/\Delta_0$ for different spin splittings (a) $h_z = 0.25$ and (b) $h_z=1.25$ both in units of $\Delta_0$, 
and (c) corresponding CPRs $J(\phi)$ in units of $e\Delta_0/(2\hbar)$. Other parameters are $k_{_S}=0$ and ${\cal M} < 0$.   
}
\label{Fig_h}
\end{figure}

Figure (\ref{Fig_h}) shows the ABS levels and CPR for two representative cases $h_z < \Delta_0$ and $h_z > \Delta_0$.
For a small spin splitting [see Fig. \ref{Fig_h}(a)], the ABS levels cross near the odd integers of $\pi$, 
experiencing also jumps of $2h_z$ when the phase passes through $2\pi N$, where $N= 1, 2, ...$. 
The level crossing is the signature of the MZMs at the adjacent ends of the right and left superconductors,
while the jumps indicate the transmutation of the spin state discussed earlier in Sec. \ref{MZMs}. 
To better understand this topological singularity let us take a closer look at the phase dependence of the ABS level $E_+(\phi)$.
When the phase approaches $2\pi$, the level evolves into a bulk state with the spin $\uparrow$. 
When the phase passes $2\pi$, the phase profile at the junction switches from a kink to an anti-kink, 
binding a bulk state with the spin $\downarrow$ which now becomes an ABS with the energy $E_+(\phi)$. 
In the presence of the field $h_z$, the energies of the two bulk spin states differ by $2h_z$, hence a finite energy jump at $2\pi$.
It is worth noting that the topological spin transmutation differs from the quantum phase transitions that occur in ferromagnetic links
when a quantum level crosses the midgap energy. \cite{Ruoco19}

As also clear from Fig. \ref{Fig_h}(a), the current is carried by the occupied (negative-energy) branches of the crossing levels
$E_-(\phi)$ and $E_+(\phi)$. The level dispersions have opposite slops, so 
the currents carried by $E_-(\phi)$ and $E_+(\phi)$ flow in the opposite directions. 
Figure \ref{Fig_h}(c) shows the corresponding CPR (for $h_z = 0.25\Delta_0$) 
with the usual sign reversal at the crossing point and the $2\pi$ periodicity 
due to the ground-state degeneracy [the occupied levels $E_\pm(\phi)$ map onto $E_\mp(\phi)$ in the next $2\pi$ phase interval].

For $h_z < \Delta_0$, the CPR (\ref{J_h}) reproduces the prediction of Ref. \onlinecite{Dolcini15} for short JJs 
within the scattering matrix approach. In fact, the calculation of the CPR in Ref. \onlinecite{Dolcini15} is more general in several respects.
One of them is the treatment of the inverse magnetoelectric effect in the superconducting leads.
It gives an extra contribution to the current, proportional to the spin splitting $h_z$ and resulting in an overall shift of the CPR. 
This paper focuses only on the ABS contribution as a probe of the topological ground states.
The omission of the bulk current is justified here, as in short JJs the phase-dependent contribution to the CPR comes mostly from the ABSs, 
and an overall shift of the CPR does not alter its shape.
 
The junction behavior changes qualitatively
as the spin splitting $h_z$ becomes equal to or exceeds the bare gap energy $\Delta_0$ [see Fig. \ref{Fig_h}(b)].
In this regime, the level crossing and the spin transmutation both occur at $2\pi N$,
with a sharp switching from one $4\pi$ -periodic ground state to the other. 
This is a realization of the topological transition discussed earlier in Sec. \ref{MZMs}. 
In Eqs. (\ref{E_h}) and (\ref{J_h}), the transitions are accounted for by the discontinuous function ${\rm sgn}[{\cal M} \sin(\phi/2)]$.
In this respect, the above results differ from the analysis of Ref. \onlinecite{Dolcini15}.

Remarkably, within each phase interval between $2\pi (N-1)$ and $2\pi N$, 
the current is carried by one fully spin-polarized ground state.
The corresponding CPR has a chiral character in the sense that the current flows in one direction independently of the applied phase bias 
[see Fig. \ref{Fig_h}(c) for $h_z = 1.25\Delta_0$]. 
The superflow direction is determined solely by the sign of the velocity of the edge channel that hosts the occupied ABS or,
in other words, by the edge-channel chirality. It is associated with the Chern number, $C$, of the occupied spin band of the 2DTI.
Indeed, for $|h_z| > \Delta_0$ Eq. (\ref{J_h}) yields
\begin{eqnarray}
J(\phi) = - \frac{e \Delta_0}{2\hbar} \, C \, 
\left|\sin\frac{\phi}{2}\right|,  
\quad 
C = {\rm sgn}({\cal M}){\rm sgn}(h_z).
\label{J_h_chi}
\end{eqnarray}
Precisely speaking, $C$ is the first Chern number of the 2DTI valence band for the antiparallel spin projection on vector ${\bm h} = [0,0,h_z]$. 
The data in Fig. \ref{Fig_h} are plotted for $h_z > 0$ (hence, the spin-down ground state) 
and inverted bulk band gap  ${\cal M} < 0$. \cite{Bernevig06, Liu16} In this case, $C = -1$.

We may inquire what would be different in non-topological JJs such as short links between 1D BCS superconductors with spin-split parabolic bands.
In the regime $h_z > \Delta_0$, there would be two occupied ABSs corresponding to a left- and a right-mover of the normal metal, each fully spin polarized. 
The currents carried by the left- and right- moving ABSs cancel each other, meaning that for $h_z > \Delta_0$ there would be no ABS contribution to the CPR.

The chiral CPR (\ref{J_h_chi}) reveals the topological transmutations of the $4\pi$ - periodic ground states 
and is, therefore, an indicator of the $4\pi$ - periodic Josephson effect. 
Still, the actual period of the CPR is $2\pi$, so the topological $4\pi$ periodicity is hidden. 
Nevertheless, the $4\pi$ periodicity shows up unmistakably in the singular ${\rm V}$ - shaped minima of $J(\phi)$, 
as indicated in Fig. \ref{Fig_h}(c). 
In fact, in order to extract the $4\pi$ periodic CPR, 
one just needs to flip the sign of $J(\phi)$ in every other $2\pi$ phase interval.
There is, however, an issue with the threshold $h_z = \Delta_0$. In practice, it requires 
strong magnetic fields or unusually large values of the g-factor. 
In the following, we seek to lower this threshold by applying bias currents or an external magnetic field.

\subsection{Case $h_z < \Delta_0$ and $k_{_S} \not =0$}
\label{h+kS}

\begin{figure}[t]
\begin{center}
\includegraphics[width=75mm]{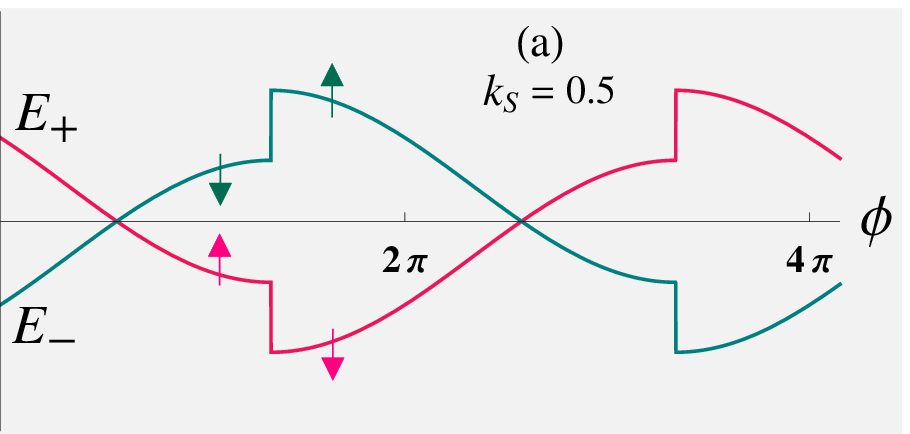}
\vskip 0.07cm
\includegraphics[width=75mm]{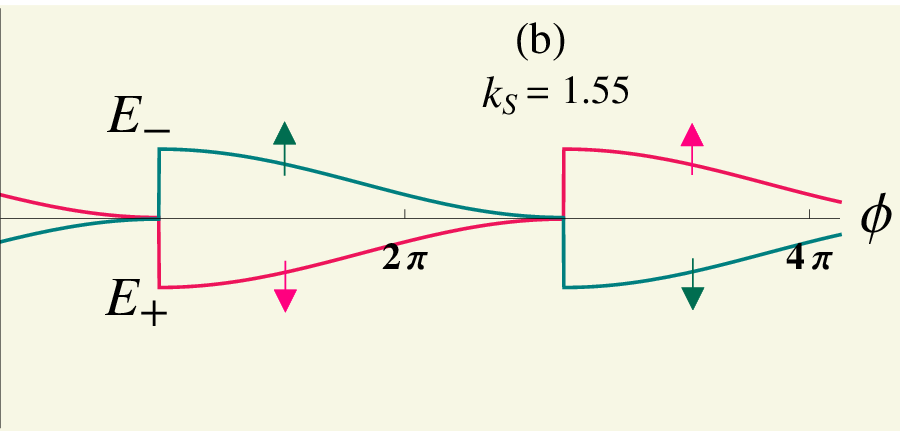}
\vskip 0.07cm
\includegraphics[width=75mm]{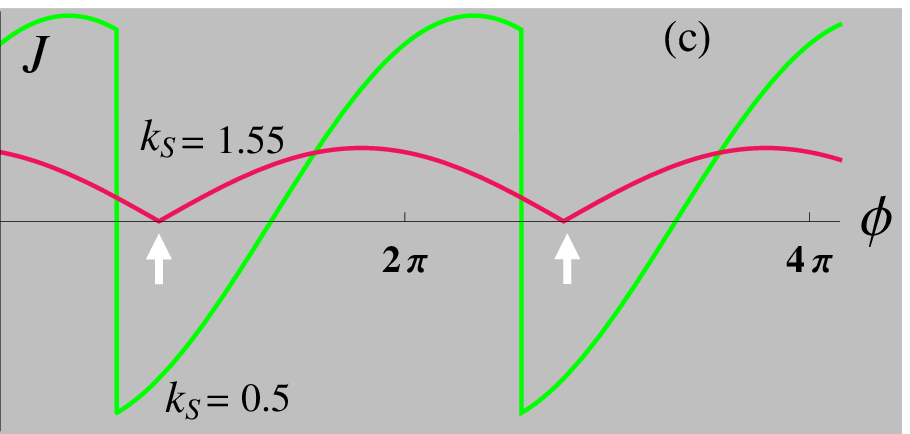}
\end{center}
\caption{
ABS levels $E_\pm(\phi)/\Delta_0$ for fixed spin splitting $h_z = 0.25 \Delta_0$ and different phase gradients 
(a) $k_{_S} = 0.5$ and (b) $k_{_S} = 1.55$ both in units of $\sqrt{\varkappa_+ \varkappa_-}$, 
and (c) corresponding CPRs $J(\phi)$ in units of $e\Delta_0/(2\hbar)$.
Other parameters are $\varkappa_- / \varkappa_+ = 0.25$ and ${\cal M} < 0$.   
}
\label{Fig_kS}
\end{figure}

Instead of increasing the spin splitting $h_z$, one can lower the pair potential by passing bias currents 
through the superconducting strips, as depicted in Fig. \ref{TJJ} [see also Ref. \onlinecite{Romito12}].
We assume that in each strip the bias current has a uniform density $j = \rho_{_S} k_{_S}$ (see also Sec. \ref{Model}). 
The dependence on $j$ translates into the $k_{_S}$ dependence of the modulus $|\Delta|$ (\ref{Delta_mod}) and the phase $\theta$ (\ref{theta}). 
The corresponding ABS levels and CPR are given by Eqs. (\ref{E_4pi}) and (\ref{J}) 
with the independent parameters $h_z$ and $k_{_S}$. 
We fix the spin splitting, choosing $h_z < \Delta_0$, and vary continuously the phase gradient.
Figure (\ref{Fig_kS}) shows the results for two representative values of the phase gradient 
$k_{_S} < \sqrt{\varkappa_+ \varkappa_-}$ and $k_{_S} > \sqrt{\varkappa_+ \varkappa_-}$.

For a small phase gradient, the main change in the behavior of the ABS levels and CPR is
the shift of the topological singularities from $2\pi N$ to $2\pi N - 2\theta(k_{_S})$ 
[see Figs. \ref{Fig_kS}(a) and (c) for $k_{_S} = 0.5 \sqrt{\varkappa_+ \varkappa_-}$].
At the same time, the gap energy $|\Delta(k_{_S})|$ (\ref{Delta_mod}) decreases,
reaching at a certain $k_{_S}$ the threshold for the full spin polarization

\begin{eqnarray}
h_z = 
|\Delta(k_{_S})| =  \frac
{\Delta_0 \varkappa_+\varkappa_-}
{\sqrt{
(\varkappa_+\varkappa_- - k_{_S}^2)^2 + (\varkappa_+ + \varkappa_-)^2 k_{_S}^2
}
}.
\label{Threshold}
\end{eqnarray}
In this regime, the current is carried by the spin-polarized ground states 
with discontinuous transitions between them at $2\pi N - 2\theta(k_{_S})$ [see Figs. \ref{Fig_kS}(b)].
The behavior resembles the case of the strong spin splitting in Fig. \ref{Fig_h}(b). 
The crucial difference is the value of $h_z$ which is 5 times smaller here.
This becomes possible because the actual threshold (\ref{Threshold}) is lower than $h_z = \Delta_0$.
Even for $h_z \ll \Delta_0$, there is a value of $k_{_S}$ at which the condition (\ref{Threshold}) is still met.
Note that proximity-induced gap $|\Delta(k_{_S})|$ does not collapse provided, of course, that 
the bias current remains below the critical value for the superconducting strips.

In the regime $|h_z| \geq |\Delta|$, the CPR (\ref{J}) transformes to 

\begin{equation}
J(\phi) = - \frac{e |\Delta|}{2\hbar} \, C \, 
\left|\sin\left(\theta + \frac{\phi}{2} \right)\right|,
\label{J_h_kS_chi}
\end{equation}
which is a generalization of Eq. (\ref{J_h_chi}) with the same Chern number $C$.
The chiral CPR (\ref{J_h_kS_chi}) is plotted in Fig. \ref{Fig_kS}(c) for $k_{_S} = 1.55\sqrt{\varkappa_+ \varkappa_-}$.
Again, the ${\rm V}$ - shaped minima of $J(\phi)$ indicate the topological transitions between the $4\pi$ - periodic ground states.

\subsection{Out-of-plane magnetic field ($h_z \not =0$ and $k_{_S} \not =0$)}
\label{B}

\begin{figure}[t]
\begin{center}
\includegraphics[width=75mm]{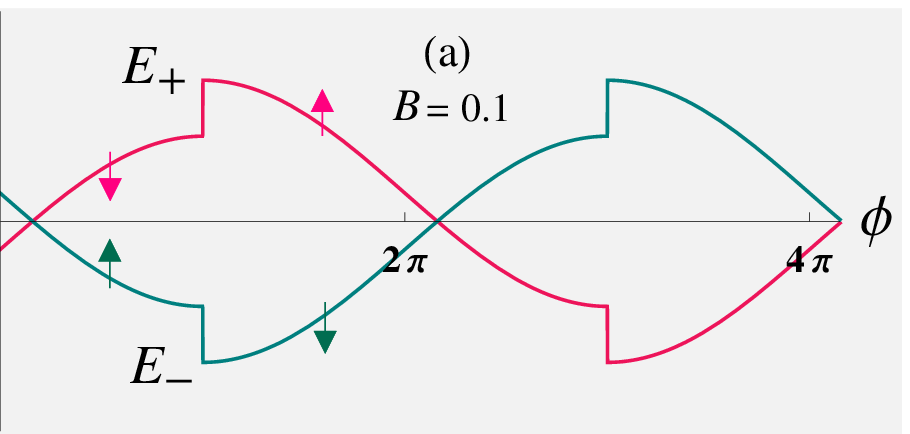}
\vskip 0.07cm
\includegraphics[width=75mm]{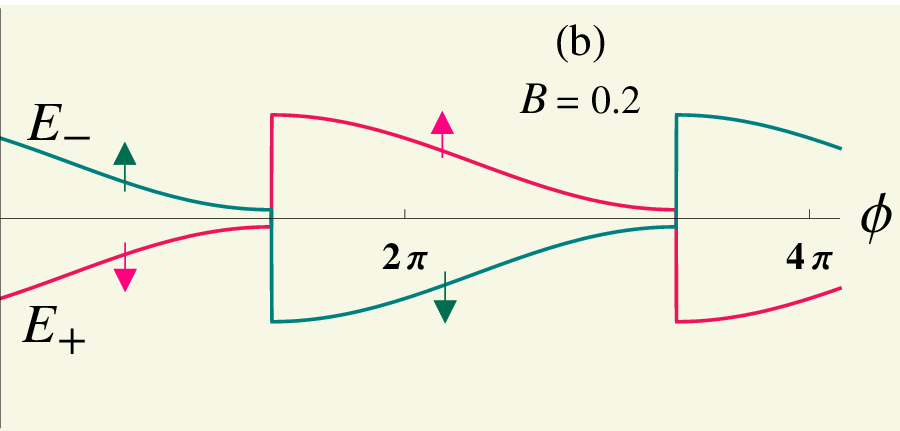}
\vskip 0.07cm
\includegraphics[width=75mm]{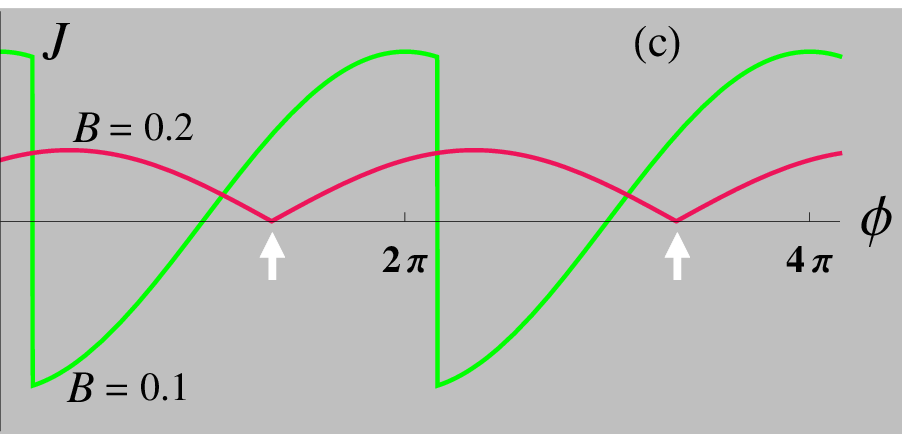}
\end{center}
\caption{
ABS levels $E_\pm(\phi)/\Delta_0$ for magnetic field strengths (a) $B = 0.1$ and (b) $B = 0.2$ both in units of $B_{spin}$,
and (c) corresponding CPRs $J(\phi)$ in units of $e\Delta_0/(2\hbar)$.
Other parameters are $B_{spin}/B_{orb} = 10$, $\varkappa_- / \varkappa_+ = 0.25$, and ${\cal M} < 0$.  
}
\label{Fig_B}
\end{figure}

We have seen that the presence of both the spin splitting and the phase gradient should allow for a less restrictive realization of the chiral CPR.
Still, the analysis above implies an independent control of $h_z$ and $k_{_S}$, which can have both advantages and disadvantages.
Perhaps the simplest situation is when $h_z$ and $k_{_S}$ are generated and controlled at once. 
This can be achieved by applying an external magnetic field perpendicularly to the 2DTI/superconductor structure. 
In this case, both $h_z$ and $k_{_S}$ can be related to the strength, $B$, of the applied field as
\begin{equation}
h_z = \frac{1}{2} g \mu_B B, \qquad k_{_S} =\frac{\pi L B}{\Phi_0}.
\label{hz_kS}
\end{equation}
Here, $h_z$ is the usual Zeeman energy, while the expression for $k_{_S}$ reflects the orbital magnetic-field effect on the superconducting contacts. \cite{GT19a} 
Above, $g$ is the Land\'e g-factor, $\mu_B$ is the Bohr magneton, $\Phi_0 = h/(2e)$ is the magnetic flux quantum, and 
$L$ is the width of one superconducting strip. Earlier, the interplay of the spin and orbital magnetic-field effects 
have been studied in connection with thermal \cite{GT04, GT05a} and electric \cite{GT05b} transport in low-dimensional semiconductor/superconductor hybrids.

The scale for the "spin" magnetic field is set by the bare energy gap $\Delta_0$:

\begin{equation}
\frac{h_z}{\Delta_0} = \frac{B}{B_{spin}}, \qquad B_{spin} = \frac{2\Delta_0}{g \mu_B}.
\label{B_spin}
\end{equation}
For a typical proximity-induced gap $\Delta_0 = 0.1$ meV  
and the electron spin g-factor $g = 2$, one has $B_{spin} \approx 1.73$ T, 
which is much higher than the critical fields of the most of superconducting materials. 
As for the "orbital" magnetic field, its scale is set by the characteristic edge-state width $(\varkappa_+\varkappa_-)^{-1/2}$: 

\begin{equation}
\frac{k_{_S}}{(\varkappa_+\varkappa_-)^{1/2}} = \frac{B}{B_{orb}},
\qquad
B_{orb} = \frac{\Phi_0}{\pi L (\varkappa_+\varkappa_-)^{-1/2}}.
\label{B_orb}  
\end{equation}
For the typical lengthscales $(\varkappa_+\varkappa_-)^{-1/2} \approx 10$ nm and $L \approx 1 \mu$m, we have  $B_{orb} \approx 0.1$ T  
(see also Ref. \onlinecite{GT19a}). 
That is, the "orbital" magnetic field is at least an order of magnitude smaller than the "spin" one.
We also assume that $B_{orb}$ is much smaller than the critical fields of the superconducting contacts, 
allowing us to disregard the pair-breaking effect in Eq. (\ref{Delta_mod}).

The magnetic-field dependence of the ABSs and CPR is obtained by inserting the expressions for the Zeeman energy (\ref{B_spin}) 
and the phase gradient (\ref{B_orb}) into Eqs. (\ref{E_4pi}) and (\ref{J}).  
As before, we focus on the phase dependence. The ratio of the "spin" and "orbital" fields is fixed to $B_{spin}/B_{orb} = 10$.
The results are shown in Fig. (\ref{Fig_B}). 
The ABS levels and CPR are clearly similar to the case of the independently controlled 
$h_z$ and $k_{_S}$. We again see the topological transitions between the two ground states [Fig. \ref{Fig_B}(b)], 
resulting in the chiral CPR  [Fig. \ref{Fig_B}(c)]. 
Most important, the threshold (\ref{Threshold}) for the full spin polarization is reached at the field value $B = 0.2 B_{spin}$ 
much smaller than the "spin" field in Eq. (\ref{B_spin}). 

In conclusion, it may be helpful to outline an experimental scheme to test the proposed theory.
The phase drop $\phi$ at the JJ can be controlled in a SQUID setup, allowing a contactless measurement of the CPR.
To achieve the spin splitting, it is necessary to apply a magnetic field at the JJ, 
permitting at the same time an independent control of $\phi$. 
According to the estimates above, the transition to the chiral CPR should occur at a rather modest field $B \approx  0.35$ T.
This sets the lower margin for the critical fields of the superconducting contacts.
This margin is sensitive to the characteristic edge-state width $(\varkappa_+\varkappa_-)^{-1/2}$.
Choosing a 2DTI material with a larger $(\varkappa_+\varkappa_-)^{-1/2}$ should further lower the threshold 
for achieving the chiral CPR. 
The estimate $B \approx  0.35$ T holds for HgTe quantum wells with superconducting Nb contacts \cite{GT19a} 
whose upper critical field is well above that value.

Finally, we may also note that the chiral CPR can be viewed as an extreme case of the directional asymmetry of the Josephson current in 
spin-orbit-coupled JJs (see, e.g., Refs. \onlinecite{Reynoso08} and \onlinecite{Zazunov09}). 
Still, what is essential here is not the directional asymmetry per se, 
but the singularity of the chiral CPR caused by the parity switching. 
This reveals two distinct ground states of the junction which always harbor a pair of MZMs.
Therefore, the chiral CPR is linked to the underlying MZMs. 
Akin to Kitaev's model, \cite{Kitaev01} each MZM is protected by the fermionic parity of the corresponding ground state.
These topological aspects distinguish the chiral Josephson effect studied here from the anomalous Josephson effect in spin-orbit-coupled JJs.
 
\acknowledgments
The author thanks J. Cayao, F. S. Bergeret and Y. Tanaka for their helpful comments.
This work was supported by the German Research Foundation (DFG) through TRR 80.

\end{document}